\documentclass[conference]{IEEEtran}

\usepackage[cmex10]{amsmath}
\usepackage{amssymb}
\usepackage[pdftex]{graphicx}

\newcommand{\ind}[1]{\mathbb{I}_{\{#1\}}}
\newcommand{\ent}{\mathcal{H}}
\newcommand{\cmap}{\psi}
\newcommand{\bigO}{{O}}
\newcommand{\risk}{R}
\newcommand{\cb}{c^\bot}
\newcommand{\Csetg}{\mathcal{C}_\gamma}
\newcommand{\dC}{\Delta\mathcal{C}}
\newcommand{\pr}{\mathbb{P}}
\newcommand{\EV}{\mathbb{E}}
\newcommand{\rate}{R}

\begin{document}

\title{
\vspace*{-10mm} {\small International Symposium on 
Information Theory 2010 (ISIT10 E-Mo-4.2), June 13-18 in Austin, TX}\\[5mm]
Information theoretic model validation for clustering
}

\author{\IEEEauthorblockN{Joachim M. Buhmann}
\IEEEauthorblockA{Department of Computer Science\\
Swiss Federal Institute of Technology, ETH Zurich\\
Email: jbuhmann@inf.ethz.ch}
}

\maketitle

\begin{abstract}
Model selection in clustering requires (i) to specify a suitable
clustering principle and (ii) to control the model order complexity by
choosing an appropriate number of clusters depending on the noise
level in the data.  We advocate an information theoretic perspective
where the uncertainty in the measurements quantizes the set of data
partitionings and, thereby, induces uncertainty in the solution space
of clusterings. A clustering model, which can tolerate a higher level
of fluctuations in the measurements than alternative models, is
considered to be superior provided that the clustering solution is
equally informative. This tradeoff between \emph{informativeness} and
\emph{robustness} is used as a model selection criterion.  The
requirement that data partitionings should generalize from one data
set to an equally probable second data set gives rise to a new notion
of structure induced information.
\end{abstract}

\IEEEpeerreviewmaketitle

%
\section{Introduction}
%

Data clustering or data partitioning has emerged as the workhorse of
\emph{exploratory data analysis}. This unsupervised learning
methodology comprises a set of data analysis techniques which group
data into clusters by either optimizing a quality criterion or by
directly employing a clustering algorithm. The zoo of models range
from centroid based algorithms like {\tt $k$-means} or {\tt
  $k$-medoids}, spectral graph methods like {\tt Normalized Cut}, {\tt
  Average Cut} or {\tt Pairwise Clustering} to linkage inspired
grouping principles like {\tt Single Linkage}, {\tt Average Linkage}
or {\tt Path-based Clustering}.

The various clustering methods and algorithms ask for a unifying
meta-principle how to choose the ``right'' clustering method dependent
on the data source. This paper advocates a shift of viewpoint away from the
problem \emph{``What is the `right' clustering model?''} to the
question \emph{``How can we algorithmically validate clustering
  models?''}. This conceptual shift roots in the assumption that
ultimately, the data should vote for their prefered model type and
model complexity\cite{Claeskens08}. Therefore, algorithms which are
endowed with the ability to validate clustering concepts can maneuver
through the space of clustering models and, dependent on the training
and validation data sets, they can select a model with maximal
information content and optimal robustness.

In this paper, we propose an information theoretic model validation
strategy to select clustering models. A clustering model is used to
generate a code for communication over a noisy channel. ``Good''
models are selected according to their robustness to noise. The
approximation precision of clustering solutions is controlled by an
algorithm called empirical risk approximation (ERA) \cite{NC-COLT}
which quantizes the hypothesis class of clusterings. ERA\ employs an
hypothetical communication framework where sets of approximate
clustering solutions for the training and for the test data are used
as a communication code. Approximations of the empirical minimizer
with model averaging over approximate solutions favors stability of
clusterings. Furthermore, it is well known that stability based model
selection \cite{NC03-LaBrRoBuh} yields highly satisfactory results in
applications although the theoretical foundation of this model
selection strategy is still controversially debated \cite{BLP:COLT06}.

%
\section{Statistical learning of clustering}
%

Given are a \textbf{set of objects} $\mathbf{O} =
\{o_1,\dots,o_n\}\in\mathcal{O}$ and measurements $\mathbf{X}\in
\mathcal{X}$ to characterize these objects. $\mathcal{O},\mathcal{X}$
denotes the object or measurement space, respectively. Such
measurements might be $d$-dimensional vectors $\mathbf{X} =
\{X_i\in\mathbb{R}^d, 1\le i\le n\}$ or relations
$\mathbf{D}=(D_{ij})\in \mathbb{R}^{n\cdot n}$ which describe the
(dis)-similarity between object $o_i$ and $o_j$. More complicated data
structures than vectors or relations, e.g., three-way data or graphs,
are used in various applications. In the following, we use the generic
notation $\mathbf{X}$ for measurements. We have to distinguish between
objects and measurements since repeated measurements might refer to
the same object. Data denote object-measurement relations
$\mathcal{O}\times \mathcal{X}$, e.g., vectorial data $\{X_i ~:~ 1\le
i\le n\}$ describe surjective relations between objects $o_i$ and
measurements $X_i := X(o_i)$.

The \textbf{hypotheses} for a clustering problem are
the functions assigning data to groups, i.e.,
\begin{eqnarray}
  c ~:~ \mathcal{O}\times \mathcal{X} &\rightarrow& \{1,\dots,k\}^n
  \nonumber\\
  (\mathbf{O},\mathbf{X}) &\mapsto& c(\mathbf{O},\mathbf{X})
\end{eqnarray}
The parameter $n=\vert \mathbf{O}\vert$ denotes the number of
objects. In cases where $\mathbf{X}$ uniquely identifies the object
set $\mathbf{O}$, i.e., there exists a bijective function between
objects and measurements, then we omit the first argument of $c$ to
simplify notation. A clustering is then denoted by $c :
\mathcal{X} \rightarrow \{1,\dots,k\}^n$.

The \textbf{hypothesis class} for a clustering problem is defined as
the set of functions assigning data to groups, i.e.,
$\mathcal{C}(\mathbf{X}) = \{c(\mathbf{O},\mathbf{X}) :
\mathbf{O}\in\mathcal{O}\}$.
For $n$ objects we can distinguish $\bigO(k^n)$ such
functions. Specific clustering models might require additional
parameters $\theta$ which characterize a cluster, e.g., the centroids in
$k$-means clustering. The hypothesis class is then the product space
of possible assignments and possible parameter values.

%
\section{Clustering costs and empirical risk approximation}
%

Exploratory pattern analysis and model selection for grouping
requires to assess the quality of clustering hypotheses. Various
criteria emphasize coherency of data or connectedness, e.g., $k$-means
clustering measures the average distance of data vectors to the
nearest cluster centroid or prototype. For the
subsequent discussion on information theoretic model validation,
a cost or risk function $\risk(c,\mathbf{X})$ is assumed to measure
how well a particular clustering with assignments $c(\mathbf{X})$
and cluster parameters $\theta$ groups the objects. To simplify the
notation, cluster parameters $\theta$ are not explicitly listed as
arguments of clustering costs but are subsumed in the specification of
the cost function $\risk$. A suitable metric for the space of
hypotheses might be chosen based on such a cost function $\risk$.

The clustering solution $\cb(\mathbf{X})$ minimizes the empirical risk
(ERM) of data clustering given the measurements $\mathbf{X}$, i.e.,
\begin{equation}
  \cb(\mathbf{X}) = \arg\min_{c} \risk(c,\mathbf{X}).
\end{equation}
Clustering solutions which are similar in costs to the ERM solution
$\cb(\mathbf{X})$ define the set $\Csetg (\mathbf{X})$ of empirical
risk approximations for clustering, i.e.,
\begin{equation}
\Csetg (\mathbf{X}) := \{c(\mathbf{X}) ~:~ \risk(c,\mathbf{X})
\le \risk(\cb,\mathbf{X}) + \gamma\}.
\end{equation}
The set $\Csetg (\mathbf{X})$ reduces to the ERM solution in the limit
$\lim_{\gamma\rightarrow 0}\Csetg (\mathbf{X}) = \{\cb(\mathbf{X})\}$.

To validate clustering methods we have to define and estimate the
generalization performance of partitionings. We adopt the two sample
set scenario with training and test data which is widely used in
statistics and statistical learning theory \cite{Vapnik82} i.e. to
bound the deviation of empirical risk from expected risk, but also for
two-terminal systems in information theory \cite{CK81}. We assume for
the subsequent discussion that training data and test data are
described by respective object sets $\mathbf{O}^{(1)},
\mathbf{O}^{(2)}$ and measurements $\mathbf{X}^{(1)}, \mathbf{X}^{(2)}
\sim \pr(\mathbf{X})$ which are drawn i.i.d. from the same probability
distribution $\pr(\mathbf{X})$. Furthermore, $\mathbf{X}^{(1)},
\mathbf{X}^{(2)}$ uniquely identify the training and test object sets
$\mathbf{O}^{(1)}, \mathbf{O}^{(2)}$ so that it is sufficient to list
$\mathbf{X}^{(j)}$ as references to object sets $\mathbf{O}^{(j)},
j=1,2$.

Statistical inference requires that clustering solutions have to
generalize from training data to test data since noise in the data
renders the ERM solution $\cb(\mathbf{X}^{(1)}) \neq
\cb(\mathbf{X}^{(2)})$ unstable. How can we evaluate the
generalization properties of clustering solutions? Before we can
evaluate the clustering costs $\risk(.,\mathbf{X}^{(2)})$ on test data
of the ERM clustering on training data $\cb(\mathbf{X}^{(1)})$ we have
to identify a clustering $c \in\mathcal{C} (\mathbf{X}^{(2)})$ which
corresponds to $\cb(\mathbf{X}^{(1)})$. A priori, it is not clear how
to compare clusterings $c(\mathbf{X}^{(1)})$ for measurements
$\mathbf{X}^{(1)}$ with clusterings $c(\mathbf{X}^{(2)})$ for
measurements $\mathbf{X}^{(2)}$. Therefore, we define the mapping
\begin{eqnarray}
\cmap : \mathcal{C}(\mathbf{X}^{(1)}) &\rightarrow&
\mathcal{C}(\mathbf{X}^{(2)}) \nonumber\\
c(\mathbf{X}^{(1)}) &\mapsto& \cmap\circ c(\mathbf{X}^{(1)})
\end{eqnarray}
which identifies a clustering hypothesis for training data
$c\in\mathcal{C}(\mathbf{X}^{(1)})$ with a clustering hypothesis for
test data $\cmap\circ c \in \mathcal{C}(\mathbf{X}^{(2)})$. The reader
should note that such a mapping $\cmap$ might change the object
indices. In cases when the measurements are elements of an underlying
metric space, then a natural choice for $\cmap$ is the nearest
neighbor mapping $\nu(i) = \arg\min_{\ell} \Vert X^{(2)}_{\ell} -
X^{(1)}_i \Vert^2$ where we identify clustering $c(\mathbf{X}^{(1)})$
with $\cmap\circ c(\mathbf{X}^{(1)}) = \left(c(X^{(2)}_{\nu(1)}),
c(X^{(2)}_{\nu(2)}), \dots,c(X^{(2)}_{\nu(n)})\right)$.

The mapping $\cmap$ enables us to evaluate clustering costs on test
data $\mathbf{X}^{(2)}$ for clusterings $c(\mathbf{X}^{(1)})$ selected
on the basis of training data $\mathbf{X}^{(1)}$. Consequently, we can
determine how many $\gamma$-optimal training solutions are also
$\gamma$-optimal on test data, i.e.,
$\Delta\mathcal{C}(\mathbf{X}^{(1)}, \mathbf{X}^{(2)}) :=
\left\vert
    \left(
        \cmap\circ\Csetg(\mathbf{X}^{(1)})
    \right)
    \cap \Csetg(\mathbf{X}^{(2)})
\right\vert$.
A large
overlap means that the training approximation set generalizes to the
test data, whereas a small or empty intersection indicates the lack of
generalization. Essentially, $\gamma$ parametrizes a coarsening of the
hypothesis class such that sets of data partitionings become stable
w.r.t measurement fluctuations. The tradeoff between stability and
informativeness is controlled by minimizing $\gamma$ under the
constraint of large $\Delta\mathcal{C}(\mathbf{X}^{(1)},
\mathbf{X}^{(2)}) / \vert \Csetg(\mathbf{X}^{(2)}) \vert$ for given
risk function $\risk(.,\mathbf{X})$.

%
\section{Coding by Approximation}
%

\begin{figure}[!t]
\centering
\includegraphics[width=\linewidth]{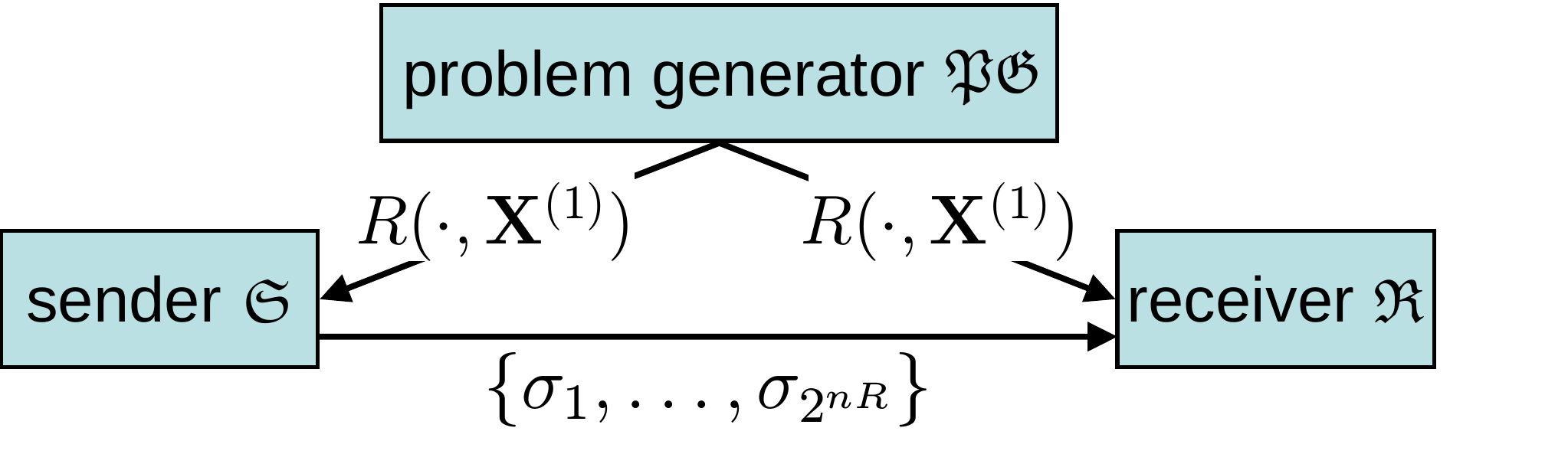}
\caption{Generation of a set of $2^{n\rate}$ code problems for
communication by e.g. permuting the object indices.}
\label{fig:1}
\end{figure}

In the following, we describe a communication scenario with a sender
$\mathfrak{S}$, a receiver $\mathfrak{R}$ and a problem generator
$\mathfrak{PG}$ where the problem generator serves as a noisy channel
between sender and receiver. Communication takes place by
approximately optimizing clustering cost functions, i.e., by
calculating approximation sets $\Csetg(\mathbf{X}^{(1)}),
\Csetg(\mathbf{X}^{(2)})$. This coding concept will be refered to as
approximation set coding (ASC). The noisy channel is characterized by
a clustering cost function $\risk(c,\mathbf{X})$ which determines the
channel capacity of the ASC scenario. Validation and selection of
clustering models is then achieved by maximizing the channel capacity
over a set of cost functions $\risk_{\theta}(.,\mathbf{X}),
\theta\in\Theta$ where $\theta$ indexes the various clustering models.

Sender $\mathfrak{S}$ and receiver $\mathfrak{R}$ agree on a
clustering principle $\risk(c,\mathbf{X}^{(1)})$ and on a mapping
function $\cmap$. The following procedure is then employed to generate
the code for the communication process:

\begin{enumerate}

\item Sender $\mathfrak{S}$ and receiver $\mathfrak{R}$ obtain a data
  set $\mathbf{X}^{(1)}$ from
  the problem generator $\mathfrak{PG}$.

\item $\mathfrak{S}$ and $\mathfrak{R}$ calculate the
  $\gamma$-approximation set $\Csetg(\mathbf{X}^{(1)})$.

\item $\mathfrak{S}$ generates a set of (random) permutations $\Sigma
  := \{\sigma_1,\dots,\sigma_{2^{n\rate}}\}$ to rename the
  objects. The permutations define a set of optimization problems
  $\risk(c,\sigma_j\circ \mathbf{X}^{(1)})$ with associated
  approximation sets $\Csetg(\sigma_j\circ \mathbf{X}^{(1)}), 1\le
  j\le 2^{n\rate}$.

\item $\mathfrak{S}$ sends the set of permutations $\Sigma$ to
  $\mathfrak{R}$ who determines the approximation sets
  $\{\Csetg(\sigma_i\circ \mathbf{X}^{(1)})\}_{i=1}^{2^{n\rate}}$.

\end{enumerate}

The rationale behind this procedure is the following: Given the
measurements $\mathbf{X}^{(1)}$ the sender has randomly covered the
set of clusterings $\mathcal{C}(\mathbf{X}^{(1)})$ by respective
approximation sets $\{\mathcal{C}(\sigma_i \circ \mathbf{X}^{(1)}):\,
1\le i\le 2^{n\rate}\}$. Communication succeeds if the approximation
sets are stable under the stochastic fluctuations of the
measurements. The criterion for reliable communication is defined by
the ability of the receiver to identify a specific permutation that
has been selected by the sender. The approximation sets
$\mathcal{C}(\sigma_i \circ \mathbf{X}^{(1)})$ play the role of
codebook vectors in Shannon's theory of communication.

After this setup procedure, both sender and receiver have a list of
approximation sets available or can algorithmically determine
membership of clusterings in one of the $2^{n\rate}$ approximation sets.

\begin{figure}
\centering
\includegraphics[width=\linewidth]{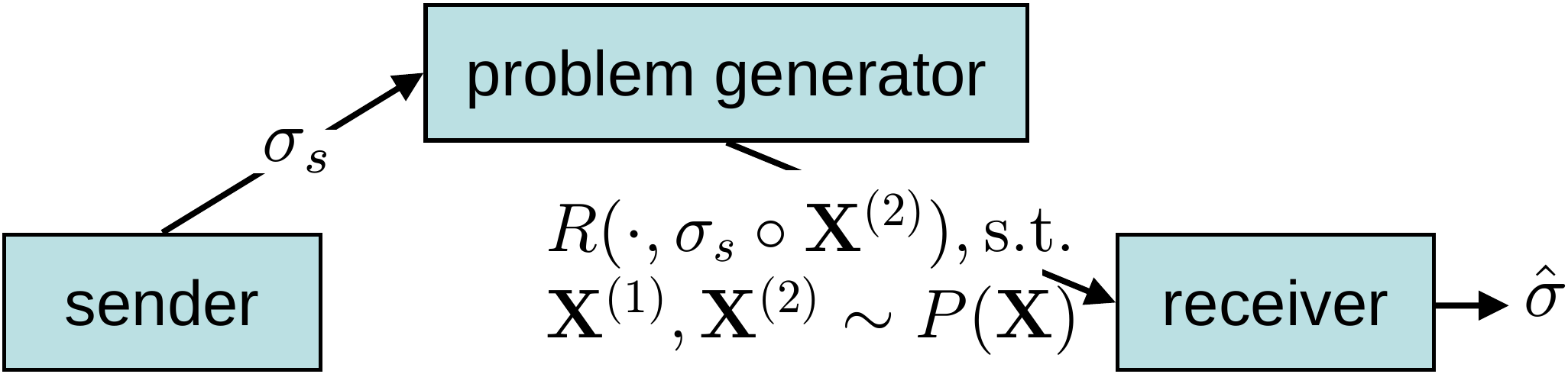}
\caption{Communication process: (1) the sender selects transformation
  $\sigma_s$, (2) the problem generator draws $\mathbf{X}^{(2)}\sim
  \pr(\mathbf{X})$ and applies $\sigma_s$ to it, and the receiver
  estimates $\sigma^\star$ based on $\tilde{\mathbf{X}} =
  \sigma_s\circ \mathbf{X}^{(2)}$.}
\label{fig:2}
\end{figure}

How is the communication between sender and receiver organized?
During communication, the following steps take place as depicted in
fig. \ref{fig:2}:
\begin{enumerate}

\item The sender $\mathfrak{S}$ selects a permutation $\sigma_s$ as
  message and send it to the problem generator $\mathfrak{PG}$.

\item $\mathfrak{PG}$ generates a new data set
  $\mathbf{X}^{(2)}$ and it applies the selected permutation to
  $\mathbf{X}^{(2)}$, yielding $\tilde{\mathbf{X}} =
  \sigma_s\circ\mathbf{X}^{(2)}$.

\item $\mathfrak{PG}$ send $\tilde{\mathbf{X}}$ to the receiver
  $\mathfrak{R}$ without revealing $\sigma_s$.

\item $\mathfrak{R}$ calculates the approximation set
  $\Csetg(\tilde{\mathbf{X}})$

\item $\mathfrak{R}$ estimates the applied permutation $\sigma_s$ by
  using the decoding rule
\begin{equation}
  \hat{\sigma} = \arg\max_{\sigma\in\Sigma} \left\vert
    \left( \cmap\circ \Csetg(\sigma\circ \mathbf{X}^{(1)}) \right) \cap
    \Csetg(\tilde{\mathbf{X}})
  \right\vert
\end{equation}
\end{enumerate}

This communication channel supports to communicate at most $n\log k$
nats if two conditions hold: (i) the channel is noise free
$\mathbf{X}^{(1)}\equiv\mathbf{X}^{(2)}$; (ii) all clusters have the
same number of objects assigned to.

It is worth mentioning that ASC is conceptually not restricted to
clustering problems although we focus the discussion here to this
problem domain.

%
\section{Error analysis of Approximation Set Coding}
%

To determine the optimal approximation precision for an optimization
problem $\risk(.,\mathbf{X})$ we have to determine necessary and
sufficient conditions which have to hold in order to reliably identify
approximation sets. Reliable identification of approximation sets
enable us to define a communication protocol using the above described
coding scheme. Therefore, we analyse the error probability of
\emph{approximation set coding} and the channel capacity which is
associated with a particular cost function $\risk(.,\mathbf{X})$. This
channel capacity will be refered to as \emph{approximation capacity}
since it determines the approximation precision of the coding scheme.

A communication error occurs if the sender selects $\sigma_s$ and the
receiver decodes $\hat{\sigma} = \sigma_j, j\neq s$. To estimate the
probability of this event, we introduce the sets
\begin{equation}
\dC_j :=
    \left(
        \cmap \circ \Csetg (\sigma_j \circ \mathbf{X}^{(1)})
    \right)
    \cap \Csetg (\tilde{\mathbf{X}}^{(2)}),~\sigma_j\in\Sigma.
\end{equation}
The set $\dC_j$ measures the intersection between the approximation set
$\Csetg (\sigma_j \circ \mathbf{X}^{(1)})$ for $\sigma_j$-permuted
measurements and the approximation set which has
been calculated by the receiver based on the test data
$\tilde{\mathbf{X}}$.

The probability of a communication error is given by a substantial
overlap $\dC_j $ with
$\sigma_j\in\Sigma\setminus\{\sigma_s\}$, i.e.,
\begin{eqnarray}
\label{proberr}
\pr(\hat{\sigma} \neq \sigma_s \vert \sigma_s)
&=&
    \pr\left(\left.
        \max_{\sigma_j\in\Sigma\setminus\{\sigma_s\}}
        \vert \dC_j \vert \ge \vert \dC_s \vert
    \right\vert \sigma_s\right) \nonumber\\
&\le& \sum_{\sigma_j\in\Sigma\setminus\{\sigma_s\}}
    \pr\left(
         \vert \dC_j \vert \ge \vert \dC_s \vert
    \bigl\vert \sigma_s\right) \\
&=& \sum_{\sigma_j\in\Sigma\setminus\{\sigma_s\}}
\EV_{\mathbf{X}^{(1,2)}} \EV_{\sigma_j}\left[\left.
    \ind{\vert \dC_j \vert \ge \vert \dC_s \vert}
\right\vert\sigma_s\right] \nonumber
\end{eqnarray}
The notation $\mathbf{X}^{(1,2)} =
(\mathbf{X}^{(1)},\mathbf{X}^{(2)})$ and
$\ind{f}={\small\begin{cases}1 & f \text{ is true} \\ 0 &
\text{otherwise} \end{cases}}$
is used.
The inequality in
(\ref{proberr}) is caused by the union bound. The confusion
probability with message $\sigma_j,j\neq s$ for given training data
$\mathbf{X}^{(1)}$ and test data $\mathbf{X}^{(2)}$ conditioned on
$\sigma_s$ is defined by
\begin{eqnarray}
\EV_{\sigma_j}\left[
    \ind{\vert \dC_j \vert \ge \vert \dC_s \vert}
\right]
&=&
\frac{1}{\vert\{\sigma_j\}\vert}
\sum_{\{\sigma_j\}}
    \ind{
        \log\vert \dC_j \vert \ge \log\vert \dC_s
        \vert
    } \nonumber\\
&\stackrel{(a)}{\le}&
\sum_{\{\sigma_j\}}
    \frac{
        \exp\left(\log\vert \dC_j \vert - \log\vert \dC_s
          \vert\right)
    }{
        \vert\{\sigma_j\}\vert
    } \nonumber\\
&=&
\frac{1}{\vert\{\sigma_j\}\vert}
\sum_{\{\sigma_j\}}
    \frac{\vert \dC_j \vert}{\vert\dC_s\vert} \nonumber\\
&\stackrel{(b)}{=}&
\frac{
    \vert\Csetg(\mathbf{X}^{(1)})\vert
    \vert\Csetg(\mathbf{X}^{(2)})\vert
}{
    \vert\{\sigma_j\}\vert \vert\dC_s\vert
} \nonumber\\
&\stackrel{(c)}{=}&
\exp\left(
    -n \mathcal{I}_\gamma(\sigma_j,\hat{\sigma})
\right)
\label{EVsigmaj}
\end{eqnarray}
The expectation $\EV_{\sigma_j}\left[ \ind{\vert \dC_j \vert \ge \vert
    \dC_s \vert} \right]$ in derivation (\ref{EVsigmaj}) is
conditioned on $\sigma_s$ which has been omitted to increase the
readability of the formulas. The summation $\{\sigma_j\}$ is indexed
by all possible realizations of the transformation $\sigma_j$ that are
uniformly selected. (a) we have used the inequality
$\ind{x\ge0}\le\exp(x)$;  (b) averaging over a random permutation
$\sigma_j$ of object indices breaks any statistical dependence between
sender and receiver approximation sets which corresponds to the error
case in jointly typical coding \cite{Cover06}; (c) we have introduced the
mutual information between the uniform distribution of the sender
message $\sigma_j$ and the receiver message $\hat{\sigma}$
\begin{eqnarray}
\label{mInfo}
\mathcal{I}_\gamma(\sigma_j,\hat{\sigma})
&=& {\frac{1}{n}}
    \log\left(
    \frac{
        \vert\{\sigma_j\}\vert \vert\dC_s\vert
    }{
        \vert\Csetg^{(1)}\vert \vert\Csetg^{(2)}\vert
    }
    \right) \\
&=& {\frac{1}{n}}
    \left(
    \log\frac{\vert\{\sigma_j\}\vert}{\vert\Csetg^{(1)}\vert} +
    \log\frac{\vert\mathcal{C}^{(2)}\vert}{\vert\Csetg^{(2)}\vert} -
    \log\frac{\vert\mathcal{C}^{(2)}\vert}{\vert\dC_s\vert}
    \right) \nonumber
\end{eqnarray}
To compactify the formula, the following notation is introduced:
$\mathcal{C}^{(i)}:= \mathcal{C}(\mathbf{X}^{(i)}), \Csetg^{(i)} :=
\Csetg(\mathbf{X}^{(i)}), i=1,2$.  The interpretation of
eq. (\ref{mInfo}) is straightforward: The first logarithm measures the
entropy of the number of transformations which can be resolved with an
uncertainty of $\Csetg^{(1)}$ in the space of clusterings on the
sender side. The logarithm
$\log({\vert\mathcal{C}^{(2)}\vert}/{\vert\Csetg^{(2)}\vert})$
calculates the entropy of the receiver clusterings which are
quantized by ${\Csetg^{(2)}}$. The third logarithm measures the joint
entropy of $(\sigma_j, \hat{\sigma})$ which depends on the size of the
intersection $\vert\dC_s\vert = \vert \left(
    \cmap \circ \Csetg (\sigma_s \circ \mathbf{X}^{(1)})
\right) \cap \Csetg (\sigma_s \circ {\mathbf{X}}^{(2)}) \vert$.

Inserting (\ref{EVsigmaj}) into (\ref{proberr}) yields the upper bound
for the error probability
\begin{eqnarray}
\label{proberr2}
\pr(\hat{\sigma} \neq \sigma_s \vert \sigma_s)
&\le&
    \exp(n\rate\log2)
    \exp\left(
    -n \mathcal{I}_\gamma(\sigma_j,\hat{\sigma})
\right) \nonumber\\
&=& \exp(-n(\mathcal{I}_\gamma(\sigma_j,\hat{\sigma}) - \rate \log2))
\end{eqnarray}

The communication rate $n\rate \log2$ is limited by the mutual
information $\mathcal{I}_\gamma(\sigma_j,\hat{\sigma})$ for asymptotically
error-free communication.

%
\section{Information theoretical model selection}
%

The analysis of the error probability suggests the following inference
principle for model selection: the approximation precision is
controlled by $\gamma$ which has to be minimized to derive more
expressive clusterings. For large $\gamma$ the rate $R$ will be low
since we resolve the space of clusterings in only a coarse grained
fashion. For too small $\gamma$ the error probability does not vanish
which indicates confusions between $\sigma_j$ and $\sigma_s$. The
optimal $\gamma$-value is given by the smallest $\gamma$ or,
equivalently the highest approximation precision
\begin{equation}
    \gamma^\star = \arg\max_{\gamma\in[0,\infty)}
    \mathcal{I}_\gamma(\sigma,\hat{\sigma}).
\end{equation}

Another choice to be made in modeling is to select a suitable cost
function for clustering $\risk(.,\mathbf{X})$.  Let us assume that a
number of cost functions $\{\risk_1(.,\mathbf{X}),
\risk_2(.,\mathbf{X}), \dots, \risk_m(.,\mathbf{X})\}$ are
considered as candidates. The cost function to be selected is
\begin{equation}
    \label{cv_rule}
    \risk^\star(c,\mathbf{X}) = \arg\max_{1\le j\le m}
    \mathcal{I}_\gamma(\sigma(\risk_j),\hat{\sigma}(\risk_j))
\end{equation}
where both the random variables $\sigma$ and $\hat{\sigma}$ depend on
$\risk(c,\mathbf{X})$. The selection rule (\ref{cv_rule}) prefers the
model which is ``expressive'' enough to exhibit high information
content (e.g., many clusters) and, at the same time robustly resists
to noise in the data set. The bits or nats which are measured in the
ASC communication setting are context sensitive since they refer to a
hypothesis class $\mathcal{C}(\mathbf{X})$, i.e., how finely or
coarsely functions can be resolved in $\mathcal{C}$.

%
\section{Computation of the approximation capacity}
%

To estimate the mutual information $\mathcal{I}_\gamma (\sigma,
\hat{\sigma})$ computationally, we have to calculate the size of the
sets $\vert\Csetg(\mathbf{X}^{(1)})\vert,\;
\vert\Csetg(\mathbf{X}^{(2)})\vert,\; \vert\{\sigma_j\}\vert,\;
\vert\dC_s\vert$.

The cardinality $\vert\{\sigma_j\}\vert$ is determined by the type of
the empirical minimizer $\cb(\mathbf{X})$, i.e., the probabilities
$\overline{p}_\nu := \pr(\cb(\mathbf{X}^{(1)})=\nu), 1\le\nu\le k$ with
\begin{equation}
    \vert\{\sigma_j\}\vert \doteq \exp( n
    \ent(\overline{p}_1,\dots,\overline{p}_k))
\end{equation}
where $\ent(\overline{p}_1,\dots,\overline{p}_k)= - \sum_{\nu=1}^k
\overline{p}_\nu \log \overline{p}_\nu$
denotes the entropy of the type of $\cb(\mathbf{X}^{(1)})$,
($a_n \doteq b_n \Leftrightarrow \lim_{n\rightarrow\infty}
\frac{1}{n} \log \frac{a_n}{b_n}=0$).

The cardinality of the approximation sets can be estimated estimated
using concepts from statistical physics. The approximation sets
$\Csetg(\mathbf{X}^{(1,2)})$ are known as microcanonical ensembles in
statistical mechanics. Estimating their size is achieved up to
logarithmic corrections by calculating the partition function
\begin{eqnarray}
    \vert\Csetg(\mathbf{X}^{(1,2)})\vert &\doteq&
    \sum_{c\in\mathcal{C}(\mathbf{X}^{(1,2)})}  \exp(- \beta
    \risk(c,\mathbf{X}^{(1,2)})).
\end{eqnarray}
The scaling factor $\beta$, also know as inverse computational
temperature, is determined such that the
average costs of the ensemble $\Csetg(\mathbf{X}^{(1)})$ yields
$\risk(\cb, \mathbf{X}^{(1)})+\gamma$. The weights $\exp(- \beta
\risk(c,\mathbf{X}^{(1,2)}))$ are known as Boltzmann factors.

The joint entropy in the mutual information, which is related to the
intersection
\begin{eqnarray}
\vert \dC \vert &=&
\left\vert\bigl(
    \cmap \circ \Csetg (\mathbf{X}^{(1)})
\bigr) \cap \Csetg (\mathbf{X}^{(2)})\right\vert \nonumber\\
&=&
\sum_{c\in\mathcal{C}(\mathbf{X}^{(2)})}
    \ind{c\in \cmap\circ\Csetg (\mathbf{X}^{(1)})}
    \ind{c\in \Csetg(\mathbf{X}^{(2)})} \nonumber\\
&\doteq&
\sum_{c\in\mathcal{C}(\mathbf{X}^{(2)})}
     \exp(- \beta \risk(\cmap^{-1}\circ c,\mathbf{X}^{(1)})) \cdot
     \nonumber\\
&&\phantom{\sum_{c\in\mathcal{C}(\mathbf{X}^{(2)})}}
     \exp(- \beta \risk(c,\mathbf{X}^{(2)})),
\end{eqnarray}
involves a product of Boltzmann factors.

The identification of approximation sets with microcanonical ensembles
provides access to a rich source of computational and analytical
methods from statistical physics to calculate the mutual information
$\mathcal{I}_\gamma (\sigma, \hat{\sigma})$. This analogy is by no
means accidental since information theory and statistical mechanics
are both specializations of empirical process theory with large
deviation analysis of many particle systems. The central role of
entropy and free energy is reflected in ASC coding where the logarithm
of the partition function arises in the mutual information
(\ref{mInfo}) twice.

The cardinalities of the approximation sets can also be numerically
estimated by sampling using Markov Chain Monte Carlo methods or by
employing analytical techniques like deterministic annealing
\cite{Rose92,IEEE-IT93}.

%
\section{Why information theory for clustering validation?}
%

There exists a long history of information theoretic approaches to
model selection, which traces back at least to Akaike's extension of
the Maximum Likelihood principle. AIC penalizes fitted models by twice
the number of free parameters. The Bayesian Information Criterion
(BIC) suggests a stronger penalty than AIC, i.e., number of model
parameters times logarithm of the number of samples. Rissanen's
minimum description length principles is closely related to BIC (see
e.g. \cite{TH:RT:JF:08} for model selection penalties). Tishby et al
\cite{NT:FP:WB:99:Allerton} proposed to select the number of clusters
according to a difference of mutual informations which is closely
related to rate distortion theory with side information.

All these information criteria regularize model estimation of the data
source. Approximation set coding pursues a different strategy for the
following reason: Quite often the measurement space $\mathcal{X}$ has
a much higher ``dimension'' than the solution space. Consider for
example the problem of spectral clustering with $k$ groups based on
dissimilarities $\mathbf{D}$: The measurements are elements of
$\mathbb{R}^{n(n-1)/2}$ for real valued, symmetric weights with
vanishing self-dissimilarities, but we can at most distinguish
$\bigO(k^n)$ different clusterings. Any approach which relies on
estimating the probability distribution $\pr(\mathbf{X})$ of the data
ultimately will fail since we require far too many observations than
needed to identify one hypothesis or a set of hypotheses, i.e., one
clustering or a set of clusterings.

Using an information theoretic perspective, we might ask the
question how the uncertainty in the measurements reduces the
resolution in the hypothesis class.
How similar can two hypotheses be so that they are still statistically
distinguishable given a cost function $\risk(c,\mathbf{X})$?
This research program is based on the idea that approximation
sets of clustering cost functions can be used as a reliable code.
The capacity of such a coding scheme then answers the question how
sensitive a particular cost function is to data noise.

%
\section{Conclusion}
%

Model selection and validation requires to estimate the generalization
ability of models from training to test data. ``Good'' models show a
high expressiveness and they are robust w.r.t. noise in the data. This
tradeoff between \emph{informativeness} and \emph{robustness} ranks
different models when they are tested on new data and it
quantitatively describes the underfitting/overfitting dilemma. In this
paper we have explored the idea to use approximation sets of
clustering solutions as a communication code. Since clustering
solutions with $k$ clusters can be represented as strings of $n$
symbols with a $k$-ary alphabet, the significant problem of model
order selection in clustering can be naturally phrased as a
communication problem. The \emph{approximation capacity} of a cost
function provides a selection criterion which renders various models
comparable in terms of their respective bit rates. The number of
reliably extractable bits of a clustering cost function
$\risk(.,\mathbf{X})$ define a ``task sensitive information measure''
since it only accounts for the fluctuations in the data $\mathbf{X}$
which actually have an influence on identifying an individual
clustering solution or a set of clustering solutions.

The maximum entropy inference principle suggests that we should
average over the statistically indistinguishible solutions in the
optimal approximation set
$\mathcal{C}_{\gamma\star}(\mathbf{X})$. Such a model averaging
strategy replaces the original cost function with the free energy and,
thereby, it defines a continuation methods with maximal
robustness. The urgent question in many data analysis applications,
which regularization term should be used without introducing an
unwanted bias, is naturally answered by the entropy. The second
question, how the regularization parameter should be selected, in
answered by ASC: Choose the parameter value which maximizes the
approximation capacity!

ASC for model selection can be applied to all combinatorial or
continuous optimization problems which depend on noisy data. The noise
level is characterized by two samples $\mathbf{X}^{(1)},
\mathbf{X}^{(2)}$. Two samples provide by far too little information
to estimate the probability density of the measurements but two large
samples contain sufficient information to determine the uncertainty in
the solution space. The equivalence of ensemble averages and time
averages of ergodic systems is heavily exploited in statistical
mechanics and it also enables us in this paper to derive a model
selection strategy based on two samples.

Future work also includes the study of
algorithmic complexity issues. The question how hard are properly
regularized optimization problems hints at a relationship between
computational complexity and statistical complexity.

\section*{Acknowledgment}
The author appreciated valuable and insightful discussions with
S. Ben-David, T. Lange, F. Pla, V. Roth and N. Tishby. This work has
been partially supported by the DFG-SNF research cluster FOR916 and by
the FP7 EU project SIMBAD.


\end{document}